\documentclass[prl,preprint,showpacs]{revtex4}

\usepackage[english]{babel}
\usepackage{graphicx}
\usepackage{bm}

\begin{document}

\title{Experimental Nonlocality Proof of
 Quantum Teleportation and Entanglement Swapping}

\author{Thomas Jennewein}
\author{Gregor Weihs}
\author{Jian-Wei Pan}
\author{Anton Zeilinger}
\affiliation{Institut f\"{u}r Experimentalphysik, Universit\"{a}t Wien
Boltzmanngasse 5, 1090 Wien, Austria}

\date{\today}

\begin{abstract}
Quantum teleportation  strikingly underlines the peculiar features of the
quantum world. We present an experimental proof of its quantum nature,
teleporting an entangled photon with such high quality that the nonlocal
quantum correlations with its original partner photon are preserved. This
procedure is also known as entanglement swapping. The nonlocality is confirmed
by observing a violation of Bell's inequality by $4.5$ standard deviations.
Thus, by demonstrating quantum nonlocality for photons that never interacted
our results directly confirm the quantum nature of teleportation.
\end{abstract}


\maketitle

Quantum state teleportation \cite{Bennett93a} allows the transfer of the
quantum state from one system to another distant one. This system becomes the
new original as it carries all information the original did and the state of
the initial particle is erased, as necessitated by the quantum no-cloning
theorem \cite{Buzek96a}. This is achieved via a combination of an entangled
state and a classical message.

The most interesting case of quantum teleportation occurs when the teleported
state itself is entangled. There the system to be teleported does not even
enjoy its own state. This procedure is also known as "Entanglement Swapping"
\cite{Zukowski93a} because (fig. 1) one starts with two pairs of entangled
photons 0--1 and 2--3, subjects photons 1 and 2 to a Bell-state measurement by
which photons~0 and 3 also become entangled. As suggested by Peres
\cite{Peres00a} this even holds if the ``entangling" Bell-state measurement is
performed after photons~0 and 3 have already been registered. Entanglement
swapping was shown \cite{Pan98a} in a previous experiment, yet the low
photon-pair visibility prevented a violation of a Bell's inequality
\cite{Bell64a} for photons~0 and 3, which is a definitive test. This is the
case, because if significant information about the state of the teleported
photon~1 were gained in the teleportation procedure, the measurements on
photons~0 and 3 would not violate Bell's inequality. This fact is
substantiated by the quantum no-cloning theorem \cite{Buzek96a}. Therefore,
the violation of Bell's inequality confirms that the state of photon~1 was
even undefined in a fundamental way and Alice could not have played any kind
of tricks to make the results look like successful teleportation. The
experiment presented here provides now such a definitive proof of the quantum
nature of teleportation.

In the present work quantum state teleportation is implemented in terms of
polarization states of photons, and hence relies on the entanglement of the
polarization of photon pairs prepared in one of the four Bell states, e.g.
\begin{equation}
|\Psi^-_{01}\rangle = \frac{1}{\sqrt{2}}(|H\rangle_0 |V\rangle_1 - |V\rangle_0
|H\rangle_1). \label{bellstate}
\end{equation}
A schematic overview of our quantum teleportation scheme is given in Fig.~1.

Initially, the system is composed of two independent entangled states and can
be written in the following way:
\begin{equation}
|\Psi_{total}\rangle = |\Psi^-\rangle_{01}\otimes|\Psi^-\rangle_{23}.
\label{productstate}
\end{equation}
Including equation (\ref{bellstate}) in (\ref{productstate}) and rearranging
the resulting terms by expressing photon~1 and photon~2 in the basis of Bell
states leads to:
\begin{equation}
|\Psi_{total}\rangle = \frac{1}{2}[|\Psi^+\rangle_{03}|\Psi^+\rangle_{12} -
|\Psi^-\rangle_{03}|\Psi^-\rangle_{12} -
|\Phi^+\rangle_{03}|\Phi^+\rangle_{12} +
|\Phi^-\rangle_{03}|\Phi^-\rangle_{12}].
\end{equation}
Alice subjects photons~1 and 2 to a measurement in a Bell-state analyzer
(BSA), and if she finds them in the state $|\Psi^-\rangle_{12}$, then
photons~0 and 3 measured by Bob, will be in the entangled state
$|\Psi^-\rangle_{03}$. If Alice observes any of the other Bell-states for
photons~1 and 2, photons~0 and 3 will also be perfectly entangled
correspondingly. We stress that photons~0 and 3 will be perfectly entangled
for any result of the BSA, and therefore it is not necessary to apply a
unitary operation to the teleported photon~3 as in the standard teleportation
protocol. But it is certainly necessary for Alice to communicate to Victor her
Bell-state measurement result. This will enable him to sort Bob's data into
four subsets, each one representing the results for one of the four maximally
entangled Bell-states.

Therefore with suitable polarization measurements on photons~0 and 3, Victor
will obtain a violation of Bell's inequality and confirm successful quantum
teleportation for each of the four subsets separately. In our experiment,
Alice was restricted to only identifying the state $|\Psi^-\rangle_{12}$ due
to technical reasons. This reduction of the teleportation efficiency to 25~\%
does not influence the fidelity. Large disturbance of the fidelity would
perturb the teleported entanglement to such a degree, that a violation of
Bell's inequality could no longer be achieved. As explained elsewhere
\cite{Bouwmeester00b} teleportation efficiency measures the fraction of cases
in which the procedure is successful and the fidelity characterizes the
quality of the teleported state in the successful cases. For example, loss of
a photon in our case leads outside the two-state Hilbert space used and thus
reduces the efficiency and not the fidelity.

It has been shown that using linear optical elements the efficiency of any BSA
is limited to maximally 50~\% \cite{Calsamiglia01a}. A configuration where
photons~1 and 2 are brought to interference at a 50:50 beam splitter is able
to identify two Bell-states exactly, and the remaining two only together (
demonstrated in \cite{Mattle96a}). Particularly easy to identify is the
$|\Psi^-\rangle_{12}$ state, as only in this case the two photons can be
detected in separate outputs of the beam splitter.

The setup of our system is shown in Fig.~2. Two separate polarization
entangled photon pairs are produced via type-II down conversion
\cite{Kwiat95b} pumped by UV laser pulses at a wavelength of 394~nm, a pulse
width of $\approx 200$~fs, a repetition rate of 76~MHz, and an average power
of 370~mW. The entangled photons had a wavelength of 788~nm. The registered
event rate of photon pairs was about $2000$ per second before  the Bell-state
analyzer (Alice) ant the polarizing beam splitter (Bob). The rate of obtaining
a four-fold photon event for the teleportation was about $0.0065$ per second.
Each single correlation measurement for one setting of the polarizers lasted
$16000$ seconds. The polarization alignment of the optical fibers performed
before each measurement proved to be stable within $1^\circ$ for 24~h.

The non-deterministic nature of the photon pair production implies an equal
probability for producing two photon pairs in separate modes (one photon each
in modes 0, 1, 2, 3) or two pairs in the same mode (two photons each in modes
0 and 1 or in modes 2 and 3). The latter can lead to coincidences in Alice's
detectors behind her beam splitter. We exclude these cases by only accepting
events where Bob registers a photon each in mode~0 and mode~3. It was shown by
Zukowski \cite{Zukowski00a}, that despite these effects of the
non-deterministic photon source experiments of our kind still constitute valid
demonstrations of nonlocality in quantum teleportation.

The entanglement of the teleported state was characterized by several
correlation measurements between photon~0 and 3 to estimate the fidelity of
the entanglement. As is customary the fidelity
$F=\langle\Psi^-|\rho|\Psi^-\rangle$ measures the quality of the observed
state $\rho$ compared to the ideal quantum case $|\Psi^-\rangle$. The
experimental correlation coefficient $E_{exp}$ is related to the ideal one
$E_{QM}$ via $E_{exp}=(4F-1)/3 \cdot E_{QM}$ \cite{note}. The correlation
coefficients are defined as $E=(N_{++}-N_{+-}-N_{-+}+N_{--})/\sum N_{ij}$,
where $N_{ij}(\phi_0,\phi_3)$ are the coincidences between the $i$--channel of
the polarizer of photon~0 set at angle $\phi_0$, and  the $j$--channel of the
polarizer of photon~3 set at angle $\phi_3$. The results (Fig.~3) show the
high fidelity of the teleported entanglement.

The Clauser-Horne-Shimony-Holt (CHSH) inequality \cite{Clauser69a} is a
variant of Bell's Inequality, which overcomes the inherent limits of a lossy
system using a fair sampling hypothesis. It requires four correlation
measurements performed with different analyzer settings. The CHSH inequality
has the following form:
\begin{equation}
S=|E(\phi_0',\phi_3')-E(\phi_0',\phi_3'')|+|E(\phi_0'',\phi_3')+
E(\phi_0'',\phi_3'')|\leq2,
\end{equation}
$S$ being the ``Bell parameter'', $E(\phi_0,\phi_3)$ being the correlation
coefficient for polarization measurements where $\phi_0$ is the polarizer
setting for photon~0 and $\phi_3$ the setting for photon~3 \cite{Aspect82b}.
The quantum mechanical prediction for photon pairs in a $\Psi^-$ state is
$E^{QM}(\phi_0,\phi_3)=-\cos(2(\phi_0-\phi_3))$. The settings
$(\phi_0',\phi_3',\phi_0'',\phi_3'') = (0^\circ,
22.5^\circ,45^\circ,67.5^\circ)$ maximize $S$ to $S^{QM}=2\sqrt{2}$, which
clearly violates the limit of 2 and leads to a contradiction between local
realistic theories and quantum mechanics \cite{Bell64a}. In our experiment,
the four correlation coefficients between photon~0 and 3 gave the following
results: $E(0^\circ,22.5^\circ)=-0.628 \pm 0.046$,
$E(0^\circ,67.5^\circ)=+0.677 \pm 0.042$, $E(45^\circ,22.5^\circ)=-0.541 \pm
0.045$, and $E(45^\circ,67.5^\circ)=-0.575 \pm 0.047$. Hence, $S=2.421 \pm
0.091$ which clearly violates the classical limit of 2 by $4.6$ standard
deviations as measured by the statistical error. The differences in the
correlation coefficients come from the higher correlation fidelity for
analyzer settings closer to $0^\circ$ and $90^\circ$, as explained in Fig.~3.

The travel time from the source to the detectors was equal within 2~ns for all
photons. Both, Alice's and Bob's detectors were located next to each other,
but Alice and Bob were separated by about 2.5~m, corresponding to a luminal
signaling time of 8~ns between them. Since the time resolution of the
detectors is  $< 1$~ns, Alice's and Bob's detection events were space like
separated for all measurements.

A seemingly paradoxical situation arises --- as suggested by Peres
\cite{Peres00a} --- when Alice's Bell-state analysis is delayed long after
Bob's measurements. This seems paradoxical, because Alice's measurement
projects photons~0 and 3 into an entangled state after they have been
measured. Nevertheless, quantum mechanics predicts the same correlations.
Remarkably, Alice is even free to choose the kind of measurement she wants to
perform on photons~1 and 2. Instead of a Bell-state measurement she could also
measure the polarizations of these photons individually. Thus depending on
Alice's later measurement, Bob's earlier results either indicate that
photons~0 and 3 were entangled or photons~0 and 1 and photons~2 and 3. This
means that the physical interpretation of his results depends on Alice's later
decision.

Such a delayed-choice experiment was performed by including two 10~m optical
fiber delays for both outputs of the BSA. In this case photons~1 and 2 hit the
detectors delayed by about 50~ns. As shown in Fig.~3, the observed fidelity of
the entanglement of photon~0 and photon~3 matches the fidelity in the
non-delayed case within experimental errors. Therefore, this result indicate
that the time ordering of the detection events has no influence on the results
and strengthens the argument of A.~Peres \cite{Peres00a}: {\em this paradox
does not arise if the correctness of quantum mechanics is firmly believed}.

One might question the ``independence'' of the photons~1 and 2 which interfere
in the BSA, since all photons are produced by down conversion from one and the
same UV-laser pulse, and the photons could take on a phase coherence from the
UV laser. Note, that the UV mirror was placed $13$~cm behind the crystal,
which greatly exceeds the pump pulse width of $\sim 60$~$\mu$m. We performed a
Mach-Zehnder interference experiment of a laser on the BSA to measure the
relative phase drifts due to instabilities of the optical paths. The
statistical analysis of the temporal phase variation was done using the Allan
variance \cite{Allan87a}, which we suggest as an appropriate measure.
Accordingly, the phase drifted in a random walk behavior, accumulated a
$1\sigma$ statistical drift of one wavelength within 400~s, and had a maximum
drift of 15~wavelengths during 10~h. In a single measurement which lasted
$16000$~seconds, any (hypothetical) phase relation between the two photons
that interfered in the BSA would have been completely washed out. Therefore
the contribution of such a phase relation to the outcome of the experiments
can be ruled out.

Our work, besides definitely confirming the quantum nature of teleportation
\cite{Zukowski00b}, is an important step for future quantum communication and
quantum computation protocols. Entanglement swapping is the essential
ingredient in quantum repeaters \cite{Briegel98a}, where it can be used to
establish entanglement between observers separated by larger distances as were
possible using links with individual pairs only.

This work was supported by the Austrian Science Fund (FWF) and the ``QuComm''
IST-FET project of the European Commission.


\begin{thebibliography}{18}
\expandafter\ifx\csname natexlab\endcsname\relax\def\natexlab#1{#1}\fi
\expandafter\ifx\csname bibnamefont\endcsname\relax
  \def\bibnamefont#1{#1}\fi
\expandafter\ifx\csname bibfnamefont\endcsname\relax
  \def\bibfnamefont#1{#1}\fi
\expandafter\ifx\csname citenamefont\endcsname\relax
  \def\citenamefont#1{#1}\fi
\expandafter\ifx\csname url\endcsname\relax
  \def\url#1{\texttt{#1}}\fi
\expandafter\ifx\csname urlprefix\endcsname\relax\def\urlprefix{URL }\fi
\providecommand{\bibinfo}[2]{#2} \providecommand{\eprint}[2][]{\url{#2}}

\bibitem[{\citenamefont{Bennett et~al.}(1993)\citenamefont{Bennett, Brassard,
  Cr\'epeau, Jozsa, Peres, and K.Wootters}}]{Bennett93a}
\bibinfo{author}{\bibfnamefont{C.~H.} \bibnamefont{Bennett}},
  \bibinfo{author}{\bibfnamefont{G.}~\bibnamefont{Brassard}},
  \bibinfo{author}{\bibfnamefont{C.}~\bibnamefont{Cr\'epeau}},
  \bibinfo{author}{\bibfnamefont{R.}~\bibnamefont{Jozsa}},
  \bibinfo{author}{\bibfnamefont{A.}~\bibnamefont{Peres}}, \bibnamefont{and}
  \bibinfo{author}{\bibfnamefont{W.}~\bibnamefont{K.Wootters}},
  \bibinfo{journal}{Phys. Rev. Lett.} \textbf{\bibinfo{volume}{70}},
  \bibinfo{pages}{1895} (\bibinfo{year}{1993}).

\bibitem[{\citenamefont{Buzek and Hillery}(1996)}]{Buzek96a}
\bibinfo{author}{\bibfnamefont{V.}~\bibnamefont{Buzek}} \bibnamefont{and}
  \bibinfo{author}{\bibfnamefont{M.}~\bibnamefont{Hillery}},
  \bibinfo{journal}{Physical Review A} \textbf{\bibinfo{volume}{54}},
  \bibinfo{pages}{1844} (\bibinfo{year}{1996}).

\bibitem[{\citenamefont{\.Zukowski et~al.}(1993)\citenamefont{\.Zukowski,
  Zeilinger, Horne, and Ekert}}]{Zukowski93a}
\bibinfo{author}{\bibfnamefont{M.}~\bibnamefont{\.Zukowski}},
  \bibinfo{author}{\bibfnamefont{A.}~\bibnamefont{Zeilinger}},
  \bibinfo{author}{\bibfnamefont{M.~A.} \bibnamefont{Horne}}, \bibnamefont{and}
  \bibinfo{author}{\bibfnamefont{A.~K.} \bibnamefont{Ekert}},
  \bibinfo{journal}{Phys. Rev. Lett.} \textbf{\bibinfo{volume}{71}},
  \bibinfo{pages}{4287} (\bibinfo{year}{1993}).

\bibitem[{\citenamefont{Peres}(2000)}]{Peres00a}
\bibinfo{author}{\bibfnamefont{A.}~\bibnamefont{Peres}},
  \bibinfo{journal}{Journal of Modern Optics} \textbf{\bibinfo{volume}{47}},
  \bibinfo{pages}{139 } (\bibinfo{year}{2000}).

\bibitem[{\citenamefont{Pan et~al.}(1998)\citenamefont{Pan, Bouwmeester,
  Weinfurter, and Zeilinger}}]{Pan98a}
\bibinfo{author}{\bibfnamefont{J.-W.} \bibnamefont{Pan}},
  \bibinfo{author}{\bibfnamefont{D.}~\bibnamefont{Bouwmeester}},
  \bibinfo{author}{\bibfnamefont{H.}~\bibnamefont{Weinfurter}},
  \bibnamefont{and}
  \bibinfo{author}{\bibfnamefont{A.}~\bibnamefont{Zeilinger}},
  \bibinfo{journal}{Phys. Rev. Lett.} \textbf{\bibinfo{volume}{80}},
  \bibinfo{pages}{3891} (\bibinfo{year}{1998}).

\bibitem[{\citenamefont{Bell}(1964)}]{Bell64a}
\bibinfo{author}{\bibfnamefont{J.}~\bibnamefont{Bell}},
  \bibinfo{journal}{Physics} \textbf{\bibinfo{volume}{1}}, \bibinfo{pages}{195}
  (\bibinfo{year}{1964}).

\bibitem[{\citenamefont{Bouwmeester et~al.}(2000)\citenamefont{Bouwmeester,
  Pan, Weinfurter, and Zeilinger}}]{Bouwmeester00b}
\bibinfo{author}{\bibfnamefont{D.}~\bibnamefont{Bouwmeester}},
  \bibinfo{author}{\bibfnamefont{J.-W.} \bibnamefont{Pan}},
  \bibinfo{author}{\bibfnamefont{H.}~\bibnamefont{Weinfurter}},
  \bibnamefont{and}
  \bibinfo{author}{\bibfnamefont{A.}~\bibnamefont{Zeilinger}},
  \bibinfo{journal}{J. Mod. Opt.} \textbf{\bibinfo{volume}{47}},
  \bibinfo{pages}{279} (\bibinfo{year}{2000}).

\bibitem[{\citenamefont{Calsamiglia and L{\"u}tkenhaus}(2001)}]{Calsamiglia01a}
\bibinfo{author}{\bibfnamefont{J.}~\bibnamefont{Calsamiglia}} \bibnamefont{and}
  \bibinfo{author}{\bibfnamefont{N.}~\bibnamefont{L{\"u}tkenhaus}},
  \bibinfo{journal}{Appl. Phys.~B} \textbf{\bibinfo{volume}{72}},
  \bibinfo{pages}{67} (\bibinfo{year}{2001}).

\bibitem[{\citenamefont{Mattle et~al.}(1996)\citenamefont{Mattle, Weinfurter,
  Kwiat, and Zeilinger}}]{Mattle96a}
\bibinfo{author}{\bibfnamefont{K.}~\bibnamefont{Mattle}},
  \bibinfo{author}{\bibfnamefont{H.}~\bibnamefont{Weinfurter}},
  \bibinfo{author}{\bibfnamefont{P.~G.} \bibnamefont{Kwiat}}, \bibnamefont{and}
  \bibinfo{author}{\bibfnamefont{A.}~\bibnamefont{Zeilinger}},
  \bibinfo{journal}{Phys. Rev. Lett.} \textbf{\bibinfo{volume}{76}},
  \bibinfo{pages}{4656} (\bibinfo{year}{1996}).

\bibitem[{\citenamefont{Kwiat et~al.}(1995)\citenamefont{Kwiat, Mattle,
  Weinfurter, Zeilinger, Sergienko, and Shih}}]{Kwiat95b}
\bibinfo{author}{\bibfnamefont{P.~G.} \bibnamefont{Kwiat}},
  \bibinfo{author}{\bibfnamefont{K.}~\bibnamefont{Mattle}},
  \bibinfo{author}{\bibfnamefont{H.}~\bibnamefont{Weinfurter}},
  \bibinfo{author}{\bibfnamefont{A.}~\bibnamefont{Zeilinger}},
  \bibinfo{author}{\bibfnamefont{A.}~\bibnamefont{Sergienko}},
  \bibnamefont{and} \bibinfo{author}{\bibfnamefont{Y.}~\bibnamefont{Shih}},
  \bibinfo{journal}{Phys. Rev. Lett.} \textbf{\bibinfo{volume}{75}},
  \bibinfo{pages}{4337} (\bibinfo{year}{1995}).

\bibitem[{\citenamefont{Zukowski}(2000{\natexlab{a}})}]{Zukowski00a}
\bibinfo{author}{\bibfnamefont{M.}~\bibnamefont{Zukowski}},
  \bibinfo{journal}{Phys. Rev.~A} \textbf{\bibinfo{volume}{61}},
  \bibinfo{pages}{022109} (\bibinfo{year}{2000}{\natexlab{a}}).

\bibitem{note} Evidently for a full
characterization of $F$ one would have to perform many more measurements.
Anyhow our observed values of $F$ give a reasonable estimate.

\bibitem[{\citenamefont{Clauser et~al.}(1969)\citenamefont{Clauser, Horne,
  Shimony, and Holt}}]{Clauser69a}
\bibinfo{author}{\bibfnamefont{J.~F.} \bibnamefont{Clauser}},
  \bibinfo{author}{\bibfnamefont{M.~A.} \bibnamefont{Horne}},
  \bibinfo{author}{\bibfnamefont{A.}~\bibnamefont{Shimony}}, \bibnamefont{and}
  \bibinfo{author}{\bibfnamefont{R.~A.} \bibnamefont{Holt}},
  \bibinfo{journal}{Phys. Rev. Lett.} \textbf{\bibinfo{volume}{23}},
  \bibinfo{pages}{880} (\bibinfo{year}{1969}).

\bibitem[{\citenamefont{Aspect et~al.}(1982)\citenamefont{Aspect, Grangier, and
  Roger}}]{Aspect82b}
\bibinfo{author}{\bibfnamefont{A.}~\bibnamefont{Aspect}},
  \bibinfo{author}{\bibfnamefont{P.}~\bibnamefont{Grangier}}, \bibnamefont{and}
  \bibinfo{author}{\bibfnamefont{G.}~\bibnamefont{Roger}},
  \bibinfo{journal}{Phys. Rev. Lett.} \textbf{\bibinfo{volume}{49}},
  \bibinfo{pages}{91} (\bibinfo{year}{1982}).

\bibitem[{\citenamefont{Allan}(1987)}]{Allan87a}
\bibinfo{author}{\bibfnamefont{D.~W.} \bibnamefont{Allan}},
  \bibinfo{journal}{IEEE Transactions on Ultrasonics, Ferroelectrics and
  Frequency Control} \textbf{\bibinfo{volume}{34}}, \bibinfo{pages}{647}
  (\bibinfo{year}{1987}).

\bibitem[{\citenamefont{Zukowski}(2000{\natexlab{b}})}]{Zukowski00b}
\bibinfo{author}{\bibfnamefont{M.}~\bibnamefont{Zukowski}},
  \bibinfo{journal}{Phys. Rev.~A} \textbf{\bibinfo{volume}{62}},
  \bibinfo{pages}{032101} (\bibinfo{year}{2000}{\natexlab{b}}).

\bibitem[{\citenamefont{Briegel et~al.}(1998)\citenamefont{Briegel, D{\"u}r,
  Cirac, and Zoller}}]{Briegel98a}
\bibinfo{author}{\bibfnamefont{H.-J.} \bibnamefont{Briegel}},
  \bibinfo{author}{\bibfnamefont{W.}~\bibnamefont{D{\"u}r}},
  \bibinfo{author}{\bibfnamefont{J.~I.} \bibnamefont{Cirac}}, \bibnamefont{and}
  \bibinfo{author}{\bibfnamefont{P.}~\bibnamefont{Zoller}},
  \bibinfo{journal}{Phys. Rev. Lett.} \textbf{\bibinfo{volume}{81}},
  \bibinfo{pages}{5932} (\bibinfo{year}{1998}).

\bibitem[{\citenamefont{Zukowski et~al.}(1995)\citenamefont{Zukowski,
  Zeilinger, and Weinfurter}}]{Zukowski95a}
\bibinfo{author}{\bibfnamefont{M.}~\bibnamefont{Zukowski}},
  \bibinfo{author}{\bibfnamefont{A.}~\bibnamefont{Zeilinger}},
  \bibnamefont{and}
  \bibinfo{author}{\bibfnamefont{H.}~\bibnamefont{Weinfurter}},
  \bibinfo{journal}{Annals of the N.Y. Acad. of Sciences}
  \textbf{\bibinfo{volume}{755}}, \bibinfo{pages}{91} (\bibinfo{year}{1995}).

\end{thebibliography}

\begin{figure}[h]

\includegraphics{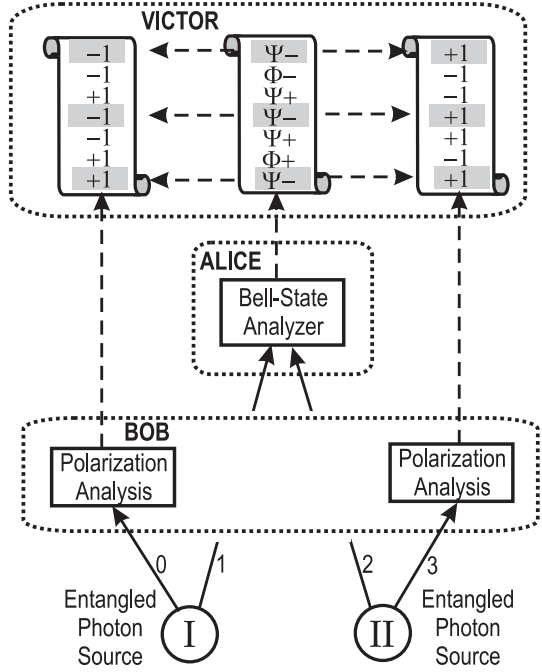}
\caption{Entanglement swapping version of quantum teleportation. Two entangled
pairs of photons 0--1 and 2--3 are produced in the sources I and II
respectively. One photon from each pair is sent to Alice who subjects them to
a Bell-state measurement, projecting them randomly into one of four possible
entangled states. Alice records the outcome and hands it to Victor. This
procedure projects photons~0 and 3 into a corresponding entangled state. Bob
performs a polarization measurement on each photon, choosing freely the
polarizer angle and recording the outcomes. He hands his results also to
Victor, who sorts them into subsets according to Alice's results, and checks
each subset for a violation of Bell's inequality. This will show whether
photons~0 and 3 became entangled although they never interacted in the past.
This procedure can be seen as teleportation either of the state of photon~1 to
photon~3 or of the state of photon~2 to photon~0. Interestingly, the quantum
prediction for the observations does not depend on the relative space-time
arrangement of Alice's and Bob's detection events.}

\end{figure}

\begin{figure}[h]
\includegraphics{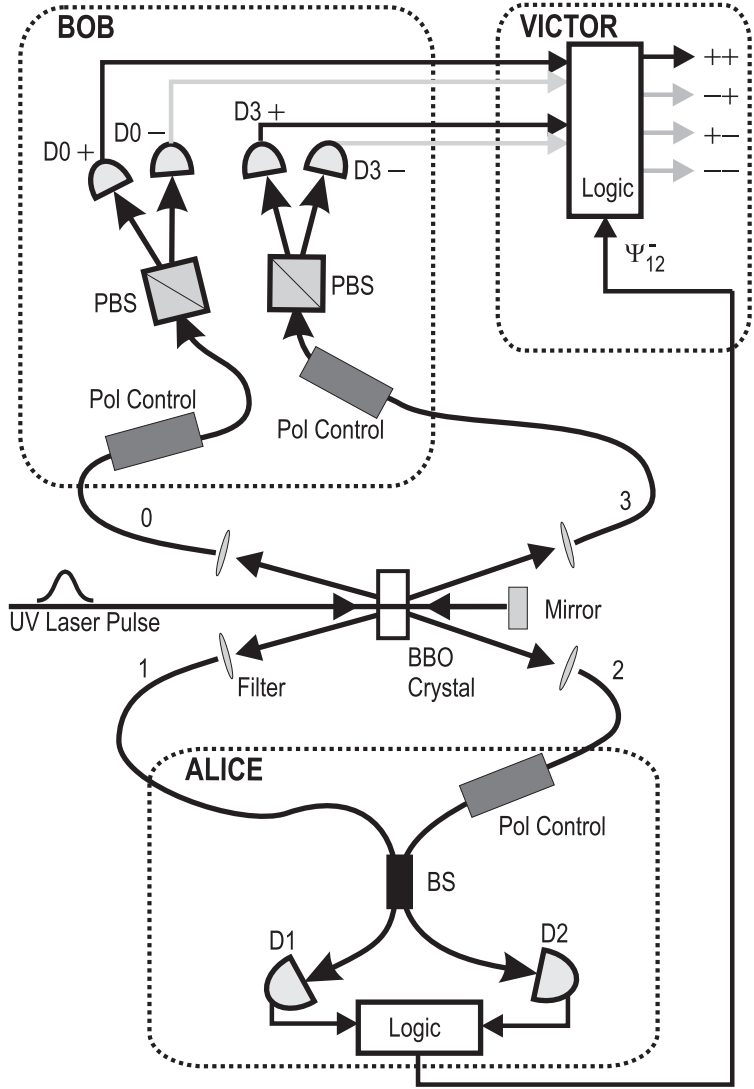}

\caption{Setup of the experiment. The two entangled photon pairs were produced
by down conversion in BBO, pumped by femtosecond UV-laser pulses traveling
through the crystal in opposite directions. Through spectral filtering with a
$\Delta \lambda_{FWHM}=3.5$~nm for photons~0 and 3 and $\Delta
\lambda_{FWHM}=1$~nm for photons~1 and 2, the coherence time of the photons
was made to exceed the pulse width of the UV-laser, making the two entangled
photon pairs indistinguishable in time, a necessary criterion for interfering
photons from independent down conversions \cite{Zukowski95a}. All photons were
collected in single-mode optical fibers for further analysis and detection.
Single-mode fibers offer the high benefit that the photons remain in a
perfectly defined spatial mode allowing high fidelity interference. For
performing the Bell-state analysis, photons~1 and 2 interfered at a fiber beam
splitter, where one arm contained a polarization controller for compensating
the polarization rotation introduced by the optical fibers. In order to
optimize the temporal overlap between photon~1 and 2 in the beam splitter, the
UV-mirror was mounted on a motorized translation stage. Photons~0 and 3 were
sent to Bob's two-channel polarizing beam splitters for analysis, and the
required orientation of the analyzers was set with polarization controllers in
each arm. All photons were detected with silicon avalanche photo diodes, with
a detection efficiency of about 40~\%. Alice's logic circuit detected
coincidences between detectors D1 and D2. It is essential, that she passes the
result as a classical signal to Victor, who determines whether Bob's detection
events violate Bell's inequality. }
\end{figure}

\begin{figure}[h]

\includegraphics{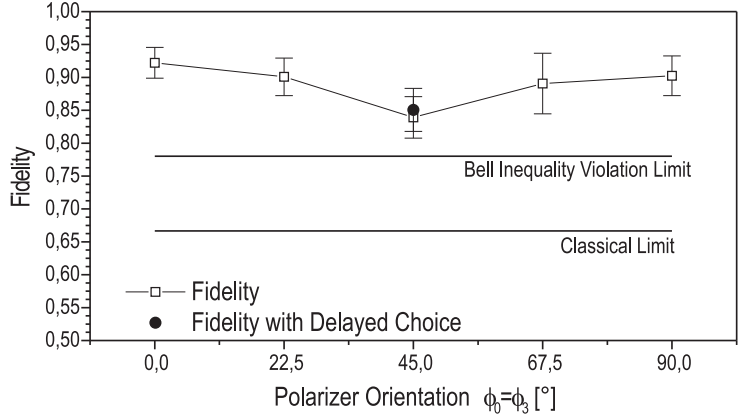}

\caption{Observed entanglement fidelity obtained through correlation
measurements between photons~0 and 3, which is a lower bound for the fidelity
of the teleportation procedure. $\phi_0$ ($\phi_3$) is the setting of the
polarization analyzer for photon~0 (photon~3) and $\phi_0=\phi_3$. The minimum
fidelity of $0.84$ is well above the classical limit of $2/3$ and also above
the limit of $0.79$ necessary for violating Bell's inequality. The fidelity is
maximal for $\phi_0=\phi_3=0^\circ,90^\circ$ since this is the original basis
in which the photon pairs are produced ($|HV\rangle$ or $|VH\rangle$). For
$\phi_0=\phi_3=45^\circ$ the two processes must interfere ($|HV\rangle -
|VH\rangle$) which is non-perfect due effects such as mismatched photon
collection or beam walk-off in the crystals. This leads to a fidelity
variation for the initially entangled pairs, which fully explains the observed
variation of the shown fidelity. Thus we conclude, that the fidelity of our
Bell-state analysis procedure is about $0.92$, independent of the
polarizations measured. The square dots represent the fidelity for the case
that Alice's and Bob's events are space-like separated, thus no classical
information transfer between Alice and Bob can influence the results. The
circular dot is the fidelity for the case, that Alice's detections are delayed
by 50~ns with respect to Bob's detections. This means, that Alice's
measurement projects photon~0 and 3 in an entangled state, at a time after
they have already been registered.}

\end{figure}

\end{document}